\documentclass[a4paper,11pt]{article}

\usepackage{jheppub}
\usepackage{graphics}%
\usepackage[mathcal]{euscript}
\usepackage[latin1]{inputenc}
\usepackage{latexsym}
\usepackage{hyperref}
\usepackage{amsmath}    
\usepackage{graphicx}   
\usepackage{verbatim}  
\usepackage{xcolor}      
\usepackage{subfigure}  
\usepackage{hyperref}   
\usepackage{bm}
\usepackage{graphicx}
\usepackage{amssymb}
\usepackage{bbm}
\usepackage{float}
\usepackage{slashed}
\usepackage{amsfonts}
\usepackage{euscript}
\usepackage{amsmath}    
\usepackage{mathrsfs} 
\usepackage{bbding}
\usepackage{pifont}
\usepackage{cancel}
\usepackage{tikz}
\usetikzlibrary{decorations.pathmorphing}
\usepackage{multirow}

\begin{document}%

\title{On the viability of regular black holes}


\author[a,b]{Ra\'ul Carballo-Rubio,}
\author[a,b]{Francesco Di Filippo,}
\author[a,b]{Stefano Liberati,}
\author[a,b]{\linebreak Costantino Pacilio}
\affiliation[a]{SISSA, International School for Advanced Studies, Via Bonomea 265, 34136 Trieste, Italy}
\affiliation[b]{INFN Sezione di Trieste, Via Valerio 2, 34127 Trieste, Italy}
\author[c]{and Matt Visser}
\affiliation[c]{School of Mathematics and Statistics, Victoria University of Wellington; PO Box 600, Wellington 6140, New Zealand}

\bigskip
\abstract{
The evaporation of black holes raises a number of conceptual issues, most of them related to the final stages of evaporation, where the interplay between the central singularity and Hawking radiation cannot be ignored. Regular models of black holes replace the central singularity with a nonsingular spacetime region, in which an effective classical geometric description is available. It has been argued that these models provide an effective, but complete, description of the evaporation of black holes at all times up to their eventual disappearance. However, here we point out that known models fail to be self-consistent: the regular core is exponentially unstable against perturbations with a finite timescale, while the evaporation time is infinite, therefore making the instability impossible to prevent. We also discuss how to overcome these difficulties, highlighting that this can be done only at the price of accepting that these models cannot be fully predictive regarding the final stages of evaporation.
}

\maketitle

\def\HRULE{{\bigskip\hrule\bigskip}}


\section{Introduction}

Black hole spacetimes in general relativity contain singularities, \emph{i.e.}, regions where the theory ceases to be predictive and no notion of classical spacetime seems to be possible. However, these singularities are hidden by event horizons, therefore being harmless for the description of physics outside black holes. This remains true, for most of their dynamical evolution, even in a semiclassical realm where black holes do evaporate due to the emission of Hawking radiation \cite{Hawking1974,Unruh1976}. It is only in the last stages of the evaporation that the existence of singularities cannot be ignored. However, these very last stages do play a very important role in determining the answer to important questions about the compatibility of general relativity and quantum mechanics --- as in the information loss problem \cite{Page1993,Mathur2009,Marolf2017,Unruh2017} --- or simply in determining the final fate of black holes. 

In this sense, a rather conservative approach is to take seriously the hypothesis that any successful theory of quantum gravity should be able to remain predictive at the would-be singular regions in general relativity \cite{Ashtekar2005,Bojowald2014}, being capable to provide at least an effective and nonsingular geometric description of the spacetime there \cite{Hayward2005,Bronnikov2006}. Regular black hole models are in this sense a useful test-bed for exploring minimal, localized, departures from classical black hole geometries and understanding if these new aspects can lead to resolution of the classical and quantum questions mentioned above. 

The nonsingular core in these geometries replaces a region in which curvature takes Planckian values or larger, and therefore in which the classical dynamics is modified \cite{Frolov1979,Frolov1981}, by a region in which curvature is bounded (this is usually known as the limiting curvature principle~\cite{Frolov1998}). In principle, it is not guaranteed that a complete quantum gravity theory allows for such an effective description, so that this is just a working assumption.

As discussed for instance in~\cite{Hayward2005,Frolov2014}, regular black holes provide an in principle complete and self-consistent picture for the evaporation of a black hole up to its disappearance. This claim is motivated by two observations: (i) the regularization of the central singularity implies the existence of both an outer and an inner horizon which, due to Hawking radiation, are naturally characterized by a dynamical evolution, wherein they become gradually closer until the black hole is extremal, and (ii) as the singular core of the evaporating black hole is regularized, there is no manifest obstruction to describing the merging and the subsequent disappearance of the two horizons in a regular manner.

Regular black holes provide therefore a framework in which long-standing problems can be analyzed without introducing dramatic deviations from standard physics. In what follows we shall analyze in more details this class of geometries, paying particular attention to two crucial aspects that were so far overlooked in the literature, instabilities and evaporation time, discussing their implication on the interpretation of regular black holes as complete and self-contained models.
In Sec. \ref{Sec:setting} we describe in generality the geometries that we will be using in the rest of the paper. In Sec. \ref{sec:inst} we show that the inner core of regular black holes is unstable. 
Sec. \ref{Sec:time} proves that the evaporation time in these models is generically infinite and discusses under which conditions this conclusion can be evaded.
Our conclusions are summarized in Sec. \ref{sec:disc}.

\section{{Geometric setting}}\label{Sec:setting}

Without loss of generality, the geometry of a spherically symmetric regular black hole can be written as \cite{Bardeen1968,Borde1996,Hayward2005,Dymnikova1992,Dymnikova2001,Ansoldi2008,DeLorenzo2014,Frolov2016,Frolov2017}
\begin{equation}\label{eq:metrictr}
\text{d}s^2=-e^{-2\phi(r)}F(r)\text{d}t^2+\frac{\text{d}r^2}{F(r)}+r^2(\text{d}\theta^2+\sin^2\theta\, \text{d}\phi^2).
\end{equation}
where $\phi(r)$ and $F(r)$ are two real functions. When convenient, we will alternatively use the notation
\begin{equation}
F(r)=1-\frac{2 m(r)}{r}.
\end{equation}
The function $m(r)$ corresponds to the Misner--Sharp--Hernandez quasi-local 
mass~\cite{Misner1964,Hernandez1966,Hayward1994}. Using the Einstein field equations, it is straightforward to see that the effective energy density associated with the geometry in Eq.~\eqref{eq:metrictr} is given by $m'(r)/4\pi r^2$, where $m'(r)=\text{d}m(r)/\text{d}r$. This quantity can be finite at $r=0$ if and only if $m(r)$ vanishes at least as $r^3$ in the limit $r\rightarrow 0$. Under the dominant energy condition, the regularity of the effective energy density also implies the regularity of the effective pressures~\cite{Dymnikova2001}. 

Hence, regularity at $r=0$ requires $F(0)=1$. On the other hand, asymptotic flatness enforces $\lim_{r\rightarrow\infty}F(r)=1$. It follows (counting the multiplicity of the roots) that the function $F(r)$ must have an even number of zeros, with these zeros corresponding to different horizons. In the following we assume for simplicity that there are only two horizons, but all our results are independent of this assumption. The outer and inner horizons, respectively $r_+$ and $r_-$, are then defined by
\begin{equation}
F(r_\pm)=0,
\end{equation}
or equivalently by
\begin{equation}
r_\pm=2m(r_\pm).
\end{equation}
When studying the causal properties of the geometry, it is convenient to introduce the ingoing Eddington-Finkelstein coordinate $v$,
\begin{equation}
\text{d}v=\text{d}t+\frac{\text{d}r}{e^{-\phi(r)}F(r)},
\end{equation}
in terms of which the line element~\eqref{eq:metrictr} reads
\begin{equation}\label{eq:metric}
\text{d}s^2=-e^{-2\phi(r)}F(r) \text{d}v^2+2e^{-\phi(r)} \text{d}r \text{d}v+ r^2(\text{d}\theta^2+\sin^2\theta\, \text{d}\phi^2).
\end{equation}
The outgoing Eddington-Finkelstein coordinate is instead given by $\text{d}u=\text{d}t-\text{d}r/e^{-\phi(r)}F(r)$. For $m(r)=M$ and $\phi(r)=0$, these expressions reduce to the standard ones in the Schwarzschild geometry.

Ingoing and outgoing radial null curves are determined respectively by the equations
\begin{equation}
\text{d}v=0,\label{eq:ingnull}
\end{equation}
and
\begin{equation}
\frac{\text{d}r}{\text{d}v}=\frac{e^{-\phi(r)}F(r)}{2}.\label{eq:outnull}
\end{equation}
The Taylor expansion of Eq.~\eqref{eq:outnull} around $r=r_\pm$ is given by
\begin{align}
{\text{d}r\over \text{d}v} &=\frac{e^{-\phi(r_\pm)}}{2}F'(r_\pm)\, (r-r_\pm) + \mathfrak{o}(r-r_\pm)\nonumber\\
&=\left. -\frac{e^{-\phi(r)}}{2}\left(2m(r)\over r\right)'\right|_{r=r_\pm}(r-r_\pm) + \mathfrak{o}(r-r_\pm);
\end{align}
we see that the surface gravities at the outer and inner horizons are given by
\begin{equation}
\label{eq:kappa:pm}
\kappa_\pm =\left.-\frac{e^{-\phi(r)}}{2}\left(2m(r)\over r\right)'\right|_{r=r_\pm}.
\end{equation}
In particular, $\kappa_+>0$ and $\kappa_-<0$. As a consequence, at the outer horizon we have an exponential peeling of outgoing null rays,
\begin{equation}
{\text{d}(r-r_+)\over \text{d}v} = |\kappa_+|\, (r-r_+) + \mathfrak{o}(r-r_H),
\end{equation}
while at the inner horizon we have an exponential focusing of outgoing null rays,
\begin{equation}
\label{eq:focus}
{\text{d}(r-r_-)\over \text{d}v} = -|\kappa_-|\, (r-r_-) + \mathfrak{o}(r-r_H).
\end{equation}
As discussed for instance in~\cite{Hayward2005,Frolov2014}, the presence of the two horizons provides in principle a complete and self-consistent picture of black hole evaporation. Indeed, when Hawking radiation is taken into account, the outer and inner horizons get closer and closer until they merge into each other: at this stage the black hole is extremal. The horizons then disappear, leaving a nonsingular remnant with a finite nonzero mass. We shall see that, in general, this picture is too simplistic and a more careful analysis is required.

\section{Instability of the inner horizon \label{sec:inst}}

The potentially unstable nature of the inner horizon due to the exponential focusing of null rays was previously noticed, and thoroughly studied, in the different but related context of charged and rotating black holes. While still being an active research area (see, for instance, the recent works~\cite{Cardoso2017,Hod2018,Dias2018,Hod:2018lmi}\footnote{In particular, it has been shown in~\cite{Cardoso2017} that the introduction of a cosmological constant may cure such instability in charged black holes. However, subsequent works have proved that this instability is still present once one takes into account charged perturbations~\cite{Hod2018}, and also on the rotating case~\cite{Dias2018,Hod:2018lmi}.}), the main aspects were settled more than two decades ago~\cite{Simpson1973,Poisson1989,Poisson1989b,Poisson1990,Barrabes1990,Ori1991}, though formal proofs of a number of technical aspects were only available later~\cite{Dafermos2003} (see also~\cite{Hamilton2008} for a review). In brief terms, the central conclusion of these works is that the inner horizon is unstable in the presence of both ingoing and outgoing perturbations, which are expected to exist in realistic collapse scenarios. While this is a classical instability, it is natural to expect the existence of semiclassical instabilities as well~\cite{Markovic1994}.

Let us start by explaining why it is natural to expect the existence of outgoing and ingoing perturbations in realistic scenarios. A regular black hole, being just a regularization of an ordinary black hole, would form after the gravitational collapse of a massive star. The star would emit radiation even after crossing the outer horizon (after this, the initially outgoing radiation is strongly lensed back to the collapsing star). On the other hand, there will be ingoing perturbations that come, for instance, from the backscattering of gravitational radiation. Interestingly, this argument has been previously applied to the classical Reissner--Nordstr\"om(-de Sitter) or Kerr(-de Sitter) black holes but, in the framework of regular black holes, it has been analyzed only for the specific case of the ``loop black hole" \cite{Brown2011}. However, here we see that it can be applied in complete generality to all regular black hole geometries.

The process can be modelled following~\cite{Barrabes1990}, where outgoing and ingoing perturbations are described in terms of null shells. This simplification allows to perform all necessary calculations analytically, and exploits the Dray--'t Hooft--Redmount (DTR) relation~\cite{Dray1985,Redmount1985} (which is also useful to describe the unstable nature of white holes~\cite{Blau1989,Barcelo2015}). 

\subsection{Mass inflation at the inner horizon}

Here, we discuss the phenomenon of mass inflation in the framework of regular black hole geometries. In this discussion, we take spherically symmetric null shells as an idealized description of these perturbations. This shell model can be understood as the discretization of what would generally be continuous streams, and therefore should preserve the main physical features. On the other hand, the approximations of spherical symmetry and null character are known to be inconsequential, in the sense that the DTR relation can be generalized to include these aspects~\cite{Barrabes1990,Nunez1993}.

\begin{figure}[h]%
\begin{center}
\vbox{
\begin{tikzpicture}[myarrow/.style={-latex},arrow in line/.style={decoration={markings,mark=at position #1 with \arrow{latex}},postaction={decorate}},dot/.style={circle,fill,inner sep=1pt}]
\node (I)    at  (0,0)  {};
\node (II)   at (-2,2)  {};
\node (III)  at (0, 4)  {};
\node (IV)   at (2,2)   {};
\node (V)  at (4, 0)    {};
\node (VI)   at (2,-2)  {};
\node (VII)  at (-2, 6) {};
\node (VIII)   at (2,6) {};

\path  
  (I)   coordinate  (I);
\path  
  (II)   coordinate  (II);
  \path  
  (III)   coordinate  (III);
\path  
  (IV)   coordinate  (IV);
\path  
  (V)   coordinate [label=0:$i^0$] (V);
\path  
  (VI)   coordinate  (VI);  
\path  
  (VII)   coordinate  (VII);
\path  
  (VIII)   coordinate  (VIII);

\draw (VI) -- (II) -- 
	node[midway, above, sloped]    {$r=r_-$}	
	(III) --
	node[midway, above, sloped]    {$r=r_-$}	
	(IV)--
	node[midway, above right]    {$\mathscr{I}^+$}
	(V) --
	node[midway, below right]    {$\mathscr{I}^-$}
	(VI);
\draw (I) --
		node[midway, below, sloped]    {$r=r_+$}
 (IV);
\draw (II) -- (VII)-- (III)-- (VIII)--(IV);


\shade (VI) to[out=120, in=-30, looseness=1] (-2,2.5)
to (II) to (VI);

\draw [line width=2.5]  (-2,2) -- (VII);

\draw[line width=2] (VIII)--(IV);
\shade (1.8,-1.8) to[out=120, in=-30, looseness=1] (-2,2.5)
to (II) to (VI);
\draw[myarrow] (-1,1) -- (0.9,2.9);
\draw[myarrow] (2,1) -- (.1,2.9);
\path  
  (0.6,2.5)   coordinate [label=0:{\small $C$}];
\path  
  (0.5,2.6)   coordinate [label=90:{\small $A$}];
\path  
  (0.4,2.5)   coordinate [label=180:{\small $D$}];
\path  
  (0.5,2.4)   coordinate [label=-90:{\small $B$}];
\fill[fill={rgb:black,1;white,1.5}] (2,-2) to[out=120, in=-30, looseness=1] (-2,2.5)
to (II) to (VI);
\end{tikzpicture}
}
\bigskip%
\caption{Schematic Penrose diagram of a star collapsing to a regular black hole with concentric outgoing and ingoing null shells. The DTR relation is applied to the crossing point $r_0$ between outgoing and ingoing shells. The corresponding four spacetime regions $A$, $B$, $C$ and $D$ are depicted.}
\label{fig:pen1}%
\end{center}
\end{figure}
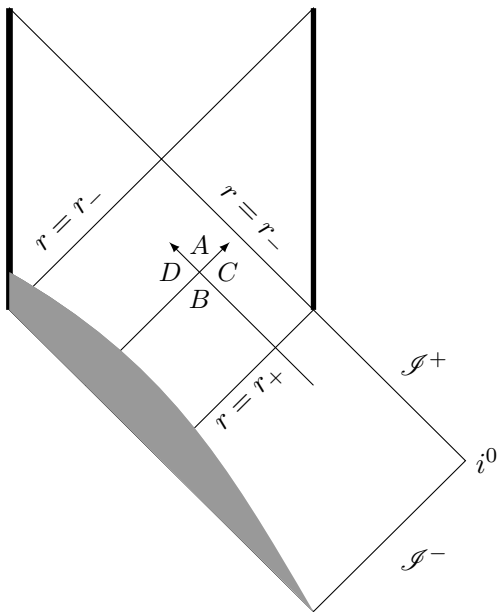

The situation we are discussing is schematically represented in the Penrose diagram of Fig. \ref{fig:pen1}. The ingoing and outgoing shells meet at the radius $r_0$ for a given moment of time. Alternatively, we can use the null coordinates $(u,v)$. In fact, we are interested in understanding the behavior of the system when this crossing point is displaced along a null outgoing curve; that is, we will take a constant value $u=u_0$ (this value is arbitrary but for the condition that it lies inside the outer horizon), and modify the value of $v$, so that the crossing point describes a curve $r_0(v)|_{u=u_0}$.

If we focus on a local neighbourhood around the crossing point $r_0$, the ingoing and outgoing shells divide the spacetime in four regions ($A$, $B$, $C$ and $D$) with different geometries and, in particular, different values of the mass parameter (in region $C$  this parameter corresponds to the Bondi mass, while in the others regions it is defined by analogy) of the corresponding regular black hole geometry. From a physical perspective, it is reasonable to expect that the mass in these four regions will be related between them and the mass of the outgoing and ingoing shells ($m_{\rm out}$ and $m_{\rm in}$, respectively), in order to satisfy certain conservation laws. This physical intuition is what the DTR relation~\cite{Dray1985,Redmount1985} makes precise. 

These relations are independent of the field equations and are formulated purely on geometric grounds. In particular, for a spherically symmetric geometry of the form~\eqref{eq:metrictr}, this relation takes the simple form~\cite{Barrabes1990}
\begin{equation}
|F_A(r_0)F_B(r_0)|=|F_C(r_0)F_D(r_0)|.\label{eq:dtr1}
\end{equation}
That is, in spherically symmetric situations, the constraint above on the coefficient $g^{rr}$ of the metric in the four spacetime regions at the crossing point must be satisfied.
Eq.~\eqref{eq:dtr1} can be manipulated in order to obtain
\begin{equation}
m_A(r_0)=m_B(r_0)+m_{\rm in}(r_0)+m_{\rm out}(r_0)-\frac{2m_{\rm out}(r_0)m_{\rm in}(r_0)}{r_0F_B(r_0)},\label{eq:dtr2}
\end{equation}
where we have $m_{\rm in}(r_0)=m_D(r_0)-m_B(r_0)$, and also $m_{\rm out}(r_0)=m_C(r_0)-m_B(r_0)$. 

The first three terms on the right-hand side of Eq. \eqref{eq:dtr2} have a clear physical meaning: $m_B$ measures the mass of the region between the ingoing and outgoing shell and, therefore, the original mass of the regular black hole before the ingoing shell is absorbed. This is moreover the region in which the coordinates $(u,v)$ are defined. On the other hand, $m_{\rm in}$ and $m_{\rm out}$ are the mass of the ingoing and outgoing shells. These three contributions are finite, but the last contribution has to be analyzed carefully. The reason is that, as the point $r_0(v)|_{u=u_0}$ gets closer to the location of the inner horizon, $F_B(r_0)\rightarrow 0$. This implies that, in order to understand the evolution of the system at late times, we need to understand the behavior with $v$ of $m_{\rm in}(r_0(v)|_{u=u_0})$ and $F_B(r_0(v)|_{u=u_0})$ (note that $m_{\rm out}$ is constant along $u=u_0$):
\begin{itemize}
\item[a)]{\emph{Behavior of $m_{\rm in}$}: this quantity describes how the ingoing tails decay with $v$, and is determined by Price's law~\cite{Price1971,Price1972,Gundlach1993,Gundlach1993b,Dafermos2003b} to be given by a power law, namely
\begin{equation}
m_{\rm in}(r_0(v)|_{u=u_0})\propto v^{-\gamma},
\end{equation}
where, for the purposes of the present discussion, it is enough to consider the lower bound $\gamma>0$.
}
\item[b)]{\emph{Behavior of $F_B$}: this quantity vanishes on the inner horizon, but we need to determine how fast it approaches this value when we displace the point $r_0$ increasing the value of $v$ along ingoing null curves. Along these trajectories Eq.~\eqref{eq:outnull} applies so that, close to the inner horizon, one has
\begin{equation}
\text{d}v=\frac{2\text{d}r}{e^{-\phi(r_-)}F'(r_-)(r-r_-)}+\mathfrak{o}(r-r_H).
\end{equation}
We just need to integrate this equation starting from some value of $r$ greater than, but arbitrarily close to, $r_-$. Taking into account that $e^{-\phi(r_-)}F'(r_-)=2\kappa_-=-2|\kappa_-|$, it follows that
\begin{equation}
F_B(r_0(v)|_{u=u_0})\propto e^{-|\kappa_-|v}.
\end{equation}
}
\end{itemize}
Combining these ingredients, we see that at late times ($v\gg 1/|\kappa_-|$), one has
\begin{equation}
m_A(r_0(v)|_{u=u_0})\propto v^{-\gamma}e^{|\kappa_-|v}.\label{eq:massinf}
\end{equation}
This is the equation that characterizes the phenomenon of mass inflation: the mass parameter in the region $A$ grows exponentially, on a timescale determined by the surface gravity of the inner horizon. In other words, the inner horizon is unstable with a characteristic timescale $1/|\kappa_-|$ measured in the ingoing null coordinate $v$. This timescale is Planckian for most of the models in the literature (see Table \ref{tab:1}). Note that the proportionality constant in Eq.~\eqref{eq:massinf} is positive, which can be realized by recalling Eq.~\eqref{eq:dtr2} and taking into account the signs of $m_{\rm out}$, $m_{\rm in}$ and $F_B(r_0)$.

In the Reissner--Nordstr\"om metric, $m_A$ is directly the physical mass in region $A$. However, in the geometries we are studying in this paper, this would not be generally the case, although this does not change the meaning of Eq.~\eqref{eq:massinf}. The reason is that $m_A$ is always proportional to the mass $M$ and, as a consequence, Eq.~\eqref{eq:massinf} can be directly translated into the unbounded growth of this parameter. In particular, on the inner horizon the function $m_A$ is generically given by a positive numerical coefficient times $M$. This will be seen explicitly in the examples discussed below.

\section{Infinite evaporation time}\label{Sec:time}

\subsection{Problem setup and working assumptions}
\label{sec:assumptions}

The instability discussed previously raises a potential inconsistency of the model. If the evaporation time is longer than the typical timescale of the instability, the backreaction of perturbations on the regular black hole geometry increases exponentially, hence modifying the geometry dramatically; it is not clear what the outcome of this dynamical evolution would be (in the classical case, the internal horizon becomes an effective spacelike singularity). To our knowledge, the evaporation time has not been studied systematically, but its finiteness is always assumed.\footnote{The only exception being the ``loop black hole''~\cite{Alesci2011}, where the evaporation is shown to happen in an infinite time, leaving no remnant.} 
In order to address the problem, we start from the most conservative viewpoint, which can be summarized in the following two  assumptions:
\begin{enumerate}
\item \emph{Adiabatic condition.} The only relevant dynamical process during the evaporation is Hawking's radiation which, at each moment of the evaporation, is thermal with temperature given by
\begin{equation}
T=\frac{\kappa_+}{2\pi}.
\end{equation}
\item \emph{Quasi-static condition.} The evaporation is a quasi-static process, meaning that the black hole passes continuously through a sequence of equilibrium states.
\end{enumerate}
We will comment on the validity of these assumptions in  Sec.~\ref{sec:peak}.

According to our hypothesis, the mass loss rate is determined by the Stefan--Boltzmann law,
\begin{equation}\label{eq:Boltz}
\frac{\text{d}M(v)}{\text{d}v}=-\sigma_{\rm SB}\,T^4(v)A_+^2(v)=-C\kappa_+^4(v)r_+^2(v),
\end{equation}
where $A_+(v)$ is the area of the outer horizon, $\sigma_{\rm SB}$ is the Stefan--Boltzman constant and $C$ is a positive constant, whose precise value is for the moment irrelevant. Here $M$, $\kappa_+$ and $r_+$ have been promoted to dynamical functions of the evaporation time $v$. Hence, from now on we will be dealing implicitly with geometries of the form
\begin{equation}\label{eq:metricv}
\text{d}s^2=-e^{-2\phi(r,M(v))}F(r,M(v)) \text{d}v^2+2e^{-\phi(r,M(v))} \text{d}r \text{d}v+ r^2(\text{d}\theta^2+\sin^2\theta\, \text{d}\phi^2).
\end{equation}
We are using the notation $\phi(r,M)$ and $F(r,M)$ in order to emphasize the dependence of these functions on the mass $M$, and also that these functions can depend on the time $v$ only implicitly, through $M(v)$.

Integration of Eq.~\eqref{eq:Boltz} requires knowledge of $\kappa_+$ and $r_+$ as functions of $M$. At the outer horizon we have
\begin{equation}\label{eq:hor}
F(r_+,M)=0.
\end{equation}
At extremality $r_+=r_\star$ and $M=M_\star$, so that the condition
\begin{equation}\label{eq:extr}
\left.\frac{\partial F(r,M_\star)}{\partial r}\right|_{r=r_\star}=0
\end{equation}
must hold. Since the surface gravity is 
\begin{equation}
\kappa_\pm=\frac{e^{-\phi\left(r_\pm\right)}}{2} \left.\frac{\partial F(r,M)}{\partial r}\right|_{r=r_\pm},
\end{equation}
the surface gravity $\kappa_\star$ of the extremal black hole vanishes, unless $\phi(r_\star,M_\star)$ diverges (this specific case is analyzed in Sec. \ref{sec:beyond}). This in turn means that the evaporation rate~\eqref{eq:Boltz} vanishes  at the end of the evaporation process. An infinitely slow evaporation rate is a clue that the evaporation time may be infinite (in fact, this has been shown to be precisely the case for near-extremal Reissner--Nordstr\"om black holes~\cite{Fabbri2000}). This would be the worst possible scenario given that, no matter how slow the mass inflation instability grows, its timescale will be trivially smaller than the evaporation time. Let us therefore explore the condition under which this happens. 

\subsection{Computation of the evaporation time}
\label{sec:time}
  
We start by assuming that the evaporation has proceeded adiabatically and quasi-statically up to a point at which the mass is arbitrarily close to the extremal value $M_\star$, $M=M_\star+\Delta M$. Our strategy then consists of integrating Eq.~\eqref{eq:Boltz} from $M_\star+\Delta M$ to $M_\star$, and showing that the corresponding time interval is infinite.

Consider a configuration in which the outer horizon radius is just an arbitrarily small $\Delta r$ away from $r_\star$
\begin{equation}
r_+=r_\star+\Delta r=r_\star\left(1+\epsilon\right),\qquad 0<\epsilon\ll 1
\end{equation}
and correspondingly the mass has an arbitrarily small deviation $\Delta M$ from $M_\star$
\begin{equation}
\label{eq:mtoext}
M=M_\star+\Delta M=M_\star\left(1+\beta\epsilon^\sigma \right)+\mathfrak{o}(\epsilon^{\sigma}),
\end{equation}
where we parametrized $\Delta M$ with two real constants $\beta$ and $\sigma>0$. These cannot be arbitrarily chosen, because $M$ and $r_+$ are related to each other by Eq.~\eqref{eq:hor}. In order to find their relation, let us expand Eq.~\eqref{eq:hor} around $\Delta r$ and $\Delta M$:
\begin{align}
\label{eq:fexp}
0=&F(r_\star,M_\star)+\left.\frac{\partial F}{\partial\,r}\right|_{r_\star,M_\star}\Delta r+\left.\frac{\partial F}{\partial\,M}\right|_{r_\star,M_\star}\Delta M+\nonumber\\
&+\left.\frac{1}{2}\frac{\partial^2 F}{\partial r^2}\right|_{r_\star,M_\star}\Delta r^2+\left.\frac{\partial^2 F}{\partial r\partial M}\right|_{r_\star,M_\star}\Delta r\,\Delta M+\nonumber\\
&\left.+\frac{1}{2}\frac{\partial^2 F}{\partial M^2}\right|_{r_\star,M_\star}\Delta M^2+\dots
\end{align}
The first term of Eq.~\eqref{eq:fexp} vanishes due to Eq.~\eqref{eq:hor} evaluated at $r_\star$ and $M_\star$, while the second term vanishes due to Eq.~\eqref{eq:extr}. Therefore, if $\partial F/\partial M\rvert_{r_\star,M_\star}\neq0$, then $\Delta M$ is \emph{at least} quadratic in $\epsilon$. Let us start considering this simpler case, and then we will analyze the most general situation. More generally, $\Delta M$ is of order $\epsilon^n$, where $n$ is the first natural number for which
\begin{equation}
\label{eq:n}
\left.\frac{\partial^n F}{\partial r^n}\right|_{r_\star,M_\star}\neq0,
\end{equation}
so that $\sigma=n$ and
\begin{equation}
\beta=-\frac{1}{n!}\left(\left.\frac{\partial F}{\partial M}\right|_{r_\star,M_\star}\right)^{-1}\left.\frac{\partial^n F}{\partial r^n}\right|_{r_\star,M_\star}.
\end{equation}
From Eq.~\eqref{eq:extr}, under the assumption that $\phi(r,M)$ is finite, we have $\kappa_\star=0$ and, therefore,
\begin{equation}
\label{eq:ktoext}
\kappa_+=\alpha\epsilon^\gamma+\mathfrak{o}(\epsilon^{\gamma}),
\end{equation}
where we have parametrized the deviation of $\kappa_+$ from $\kappa_\star=0$ with two real constants $\alpha$ and $\gamma>0$. Plugging Eqs.~\eqref{eq:mtoext} and~\eqref{eq:ktoext} in the Stefan-Boltzmann law~\eqref{eq:Boltz}, we can integrate the latter to obtain the evaporation time $\Delta v$:
\begin{equation}
\Delta v=-\frac{M_\star}{r_\star^2\,C}\frac{\beta\sigma}{\alpha^4}\int_{\epsilon_0}^{0} d\epsilon \,\epsilon^{\sigma-4\gamma-1}.
\end{equation}
We see that $\Delta v$ is finite if and only if
\begin{equation}\label{eq:fin_time}
\sigma-4\gamma>0.
\end{equation}
However,
\begin{align}
\label{eq:kexp}
\kappa_+&=\frac{e^{-\phi(r_+)}}{2}\left.\frac{\partial F}{\partial r}\right|_{r_+}\nonumber\\
&=\frac{e^{-\phi(r_+)}}{2}\sum_{i,j=0}^\infty\left.\frac{1}{i!j!}\frac{\partial^{(i+j+1)}F}{\partial^{(i+1)}r\partial^{(j)}M}\right|_{r_\star,M_\star}r_\star^i\epsilon^i\Delta M^j,
\end{align}
where in the second line we have used the Taylor expansion of $\left.\partial F/\partial r\right|_{r_+}$. The leading term in the expansion  is
\begin{equation}
\label{eq:kexp:2}
\kappa_+=\frac{e^{-\phi(r_\star)}}{2}\left.\frac{1}{n!}\frac{\partial^{n}F}{\partial r^n}\right|_{r_\star,M_\star}r_\star^{n-1}\epsilon^{n-1}+\mathfrak{o}(\epsilon^{n-1}),
\end{equation}
where $n$ is the same as in Eq.~\eqref{eq:n}. It follows that $\gamma=n-1$ and $\sigma-4\gamma=4-3n$. Given that $n\geq2$, Eq.~\eqref{eq:fin_time} cannot be satisfied. 

\begin{figure}[h]
\begin{center}
\includegraphics[scale=0.8]{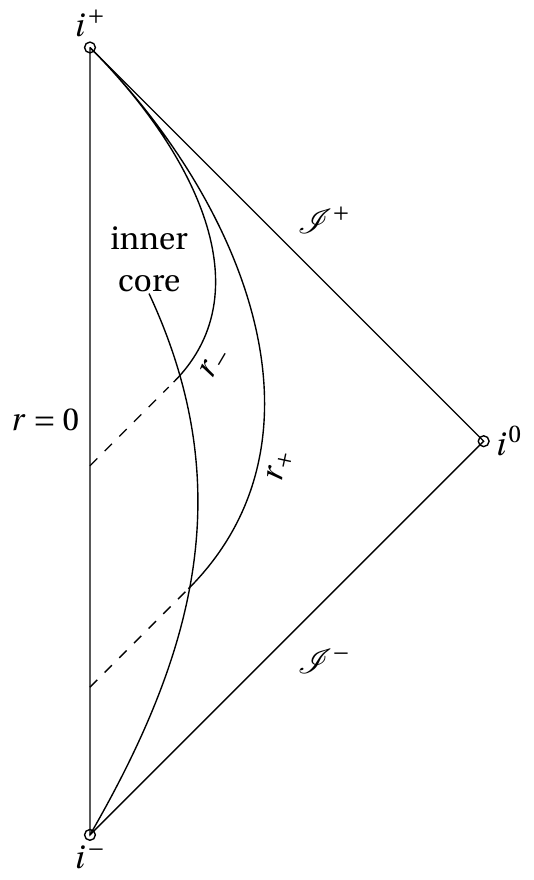}
\caption{Penrose diagram for the formation and evaporation in an infinite time of a regular black hole.}\label{fig_evaporation}
\end{center}
\end{figure}

Let us now relax the assumption $\partial F/\partial M\rvert_{r_\star,M_\star}\neq0$. Eqs.~\eqref{eq:fexp},~\eqref{eq:fin_time} and~\eqref{eq:kexp} are completely generic, and sufficient that the evaporation time is infinite. Indeed, let us distinguish three cases:
\begin{description}
\item[Case $\bm{\sigma=1}$] Eq.~\eqref{eq:fin_time} becomes $1-4\gamma>0$. But, from Eq.~\eqref{eq:kexp}, $\gamma$ must be a positive integer. Hence a contradiction.
\item[Case $\bm{\sigma<1}$] From Eq.~\eqref{eq:kexp} there are two subcases. Either $\gamma$ is a positive integer and Eq.~\eqref{eq:fin_time} is trivially violated, or $\gamma=I+J\sigma$ where $I\geq0$ and $J>0$ are appropriate integers. In the last subcase $\gamma\geq\sigma$, hence $\sigma-4\gamma<-3\sigma<0$, which contradicts Eq.~\eqref{eq:fin_time}.
\item[Case $\bm{\sigma>1}$] The term of order $\Delta r^n\sim\epsilon^n$ in Eq.~\eqref{eq:fexp}, where $n$ is defined as in Eq.~\eqref{eq:n}, can be canceled only by terms of order $\Delta r^I\Delta M^J\sim\epsilon^{I+J\sigma}$, where $I,J\geq1$ are appropriate integers such that $n=I+J\sigma$. This implies $\sigma\leq n$. On the other hand, $\gamma=n-1$ for the same reasoning of Eq.~\eqref{eq:kexp:2}. Therefore $\sigma-4\gamma\leq4-3n<0$, where the last inequality follows from $n\geq2$. Hence, Eq.~\eqref{eq:fin_time} cannot be satisfied.
\end{description}
Therefore, we conclude that the evaporation time is infinite for analytic spacetime geometries. Intuitively, the result follows from the fact that, in the extremal limit, the surface gravity goes to zero sufficiently fast and makes the evaporation rate indefinitely slow. It is straightforward to realize that a generic, but non-divergent, $\phi(r,M)$ would not change the conclusion. This infinite timescale has important implications for the global structure of the spacetime (see Fig. \ref{fig_evaporation}).

\begin{center}
\begin{table}[b]
\begin{center}
\hspace*{-.5cm}
\begin{tabular}{||c|c|c||c|c||c|c|c||}
\hline\hline
 Model & \multicolumn{2}{c}{Metric}\vline\,\vline & \multicolumn{2}{c}{Inner horizon}\vline\,\vline  & \multicolumn{3}{c}{{\begin{tabular}[c]{@{}c@{}}Evaporation \\ time\end{tabular}} }\vline\,\vline\\

\hline\hline
& $m(r)$ & $|\phi(r)|$ & $ r_-$ & $\kappa_-$ & $\sigma$ & $\gamma$ &$\Delta t$\\
\hline\hline
Hayward~\cite{Hayward2005} & $\frac{Mr^3}{(r^3+2M\ell^2)}$ & 0 & $\ell$ & $-\frac{1}{\ell}$ &\multirow{4}{*}{2} &\multirow{4}{*}{1} &\multirow{4}{*}{$\infty$} \\
\cline{1-5}
Hayward--Frolov--Zelnikov~\cite{Frolov2017} & $\frac{Mr^3}{(r^3+2M\ell^2)}$ & $<\infty$ & $\ell$ & $-\frac{e^{-\phi(r_-)}}{\ell}$ & & & \\
\cline{1-5}
Bardeen~\cite{Bardeen1968,Borde1996} & $\frac{Mr^3}{ \left[r^2+(2M\ell^2)^{2/3}\right]^{3/2}}$ & $0$ & $\sqrt{\frac{\ell^3}{2M}}$ & $-\frac{1}{\ell}$ & & & \\
\cline{1-5}
Dymnikova~\cite{Dymnikova1992,Dymnikova2001} & $M\left[ 1-e^{\left(  -r^3/2M\ell^2 \right)}  \right]$ & $0$ & $\ell$ & $-\frac{1}{\ell}$ & & & \\
\hline\hline  
\end{tabular}
\caption{The most relevant properties of the examples that we have considered. The parameter $\ell$ is a length scale, which is usually identified with the Planck length. Only the leading order of $r_-$ and $\kappa_-$ is provided, so that these quantities must be multiplied by $[1+\mathscr{O}(\ell/M)]$ terms.}
\label{tab:1}
\end{center}
\end{table}
\end{center}

Table \ref{tab:1} summarizes the properties of several models presented in the literature, and makes explicit that these models have an infinite evaporation time. It may seem surprising that, in all these models, the exponents $\sigma$ and $\gamma$ are the same. However, from Eq.~\eqref{eq:fexp} one can read that these coefficients are in general model independent, unless the function $F(r,M)$ is tuned so that some of its derivatives vanish.

\subsection{Adiabatic and quasi-static conditions}
\label{sec:peak}

In this section we discuss the validity of the two hypotheses listed in Sec.~\ref{sec:assumptions}.

\subsubsection{Adiabatic condition}

Hawking radiation is derived assuming that the geometry is static or, at the very least, evolving slow enough in a certain sense. The precise meaning of the word ``slow'' in this framework was defined in~\cite{Barcelo2010,Barcelo2010b}, where it was shown that the thermality of Hawking radiation is guaranteed if the following adiabatic condition is satisfied:
\begin{equation}
 \frac{1}{\kappa_+^2}\frac{\text{d}\kappa_+}{\text{d}v}\ll 1.
\end{equation} 
For regular black holes, this condition might only fail during the final stages of the evaporation, when both $\text{d}\kappa_+/\text{d}v$ and $\kappa_+^2$ go to zero. From Eq. \eqref{eq:ktoext}, it follows that
\begin{equation}
\frac{\text{d}\kappa_+}{\text{d}v}\propto \epsilon^{\gamma-1}\frac{\text{d}\epsilon}{\text{d}v},
\end{equation}
while taking the derivative in Eq. \eqref{eq:mtoext} and using the evaporation law \eqref{eq:Boltz} leads to
\begin{equation}
\epsilon^{\sigma-1}\frac{\text{d}\epsilon}{\text{d}v}\propto\frac{\text{d}M}{\text{d}v}\propto \epsilon^{4\gamma}.
\end{equation}
This permits to solve for $\text{d}\epsilon/\text{d}v$, so that the adiabatic condition becomes
\begin{equation}
\frac{1}{\kappa_+^2}\frac{\text{d}\kappa_+}{\text{d}v}\propto \epsilon^{3\gamma-\sigma}.
\end{equation}
Hence, it follows that the adiabatic condition is satisfied if and only if
\begin{equation}\label{eq:adiabatic}
3\gamma-\sigma>0.
\end{equation}
Repeating the same analysis performed at the end of Sec. \ref{sec:time}, one can easily show that this condition is always satisfied for spacetime geometries that are analytic.

\subsubsection{Quasi-static condition}

We have studied explicitly only the late stages of the evaporation, in which the temperature is arbitrarily small so that the quasi-static approximation is surely valid in this stage. However, we implicitly assumed that the black hole, starting from a macroscopic mass much greater than $M_\star$, reaches a mass $M_\star+\Delta M$ with $\Delta M \ll M_\star$ through an adiabatic and quasi-static process driven by the Hawking radiation.

\begin{figure}
\begin{center}
\includegraphics[scale=.6]{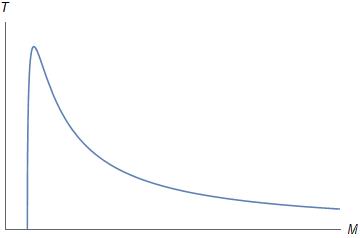}
\caption{Qualitative plot of the Hawking temperature as a function of the mass.}\label{Fig:temp}
\end{center}
\end{figure}

Since the surface gravity becomes zero both for $M=M_\star$ and at $M\to\infty$, it must have a global maximum for some value $M_\text{peak}$ of the black hole mass. If the corresponding Hawking temperature $T_\text{peak}$ is high enough, then the evaporation process can enter in a regime in which most of its mass is emitted in a short time, so that the quasi-static approximation no longer holds. Therefore, we must address whether or not this is the case. For concreteness, we refer to what is probably the simplest regular black hole model, originally proposed by Hayward~\cite{Hayward2005}. Hayward's metric is specified by
\begin{equation}
\label{eq:hay1}
F(r,M)=1-\frac{2Mr^2}{r^3+2M\ell^2},\qquad\qquad \phi(r,M)=0.
\end{equation}
The outer and inner horizons $r_\pm$ for a given mass are identified by the relation
\begin{equation}
\label{eq:rpm}
M=\frac{r_\pm^3}{2\left(r_\pm^2-\ell^2\right)}\,.
\end{equation}
When $M\to\infty$, $r_+\to2M$ and $r_-\to \ell$. The function~\eqref{eq:rpm} has a global minimum for $r_\star=\sqrt{3}\ell$ and $M_\star=3\sqrt{3}\ell/4$, which corresponds to the extremal mass. The surface gravity of the outer horizon is
\begin{equation}
\kappa_+=\frac{3}{4M}-\frac{1}{r_+}\,.
\label{eq:hayk}
\end{equation}
Fig.~\ref{Fig:temp} shows the dependence of $\kappa_+/2\pi$ on the mass $M$: We see that there is only one maximum at $M=M_\text{peak}$.
%
%
In order to find the exact value of $M_\text{peak}$ it is sufficient to minimize the function
\begin{equation}
\label{eq:lag}
\tilde{F}(r,M)=\frac{\partial F(r,M)}{\partial r}+\tilde{\lambda}F(r,M)
\end{equation}
with respect to $r$, $M$ and $\tilde{\lambda}$, finding $r_\text{peak}=3\ell$ and $M_\text{peak}=27\ell/16$. The corresponding height of the temperature peak is
\begin{equation}
T_\text{peak}=\frac{\kappa_\text{peak}}{2\pi}=\frac{1}{18\pi \ell}\approx\frac{10^{-2}}{\ell}\,.
\end{equation}
Therefore, if we identify $\ell$ with the Planck length, the peak temperature is two orders of magnitude smaller than the Planck temperature. In order to assess the extent to which this is consistent with the quasi-static condition, let us point that at $M_\text{peak}$ the system is dominated only by the scale $\ell$. Therefore we can roughly estimate the mass loss at the peak by multiplying the peak luminosity
\begin{equation}
\left.\frac{dM}{dt}\right|_\text{peak}=-\sigma_{\rm SB}\,T_\text{peak}^4 4\pi r_\text{peak}^4
\end{equation}
by the characteristic timescale at the peak, $t_\text{peak}\approx \ell$. Identifying $\ell=L_\text{planck}$ we find that the emitted mass is about $10^{-6}M_\text{peak}$.  We interpret this result as an indication that the system is not much perturbed even at the peak point, implying that the adiabatic approximation is reasonable.

We have numerically computed $r_\text{peak}$ and $M_\text{peak}$ for the ``Bardeen black hole''~\cite{Bardeen1968,Borde1996} and the ``Dymnikova black hole''~\cite{Dymnikova1992}, finding that the mass emitted at the peak is $10^{-7} M_\text{peak}$ in the former case and $10^{-5} M_\text{peak}$ in the latter one. A similar conclusion was reached in~\cite{Falls2014} in the context of asymptotic safety.

\subsection{Beyond the quasi-static or adiabatic regimes \label{sec:beyond}} 

Comparing Eqs.~(\ref{eq:fin_time}) and~(\ref{eq:adiabatic}) it is evident that it is not possible to satisfy the adiabatic condition and also have a finite evaporation time (this follows from $\sigma>0$). Therefore, the only way to build a model with finite evaporation time is going outside either the quasi-static or the adiabatic regimes.

It is possible to go outside the quasi-static approximation by simply picking a $\phi(r,M)\neq0$  that is very large at the peak mass. In this case the system would enter in a regime that requires beyond semiclassical physics to be addressed (in other words, the effective description in terms of the regular black hole geometry breaks down).

Evading the adiabatic condition is less trivial. One would need to pick a function $\phi(r,M)$ such that it diverges at $r_+=r_\star$ when $M=M_\star$. Indeed, returning to Eqs.~\eqref{eq:kexp:2} and \eqref{eq:adiabatic}, it is sufficient for $\sigma-3\gamma$ to be positive that
\begin{equation}
e^{-\phi(r_\star+\Delta r,M_\star+\Delta M)}\propto\epsilon^{-\delta},
\end{equation}
where
\begin{equation}
\label{eq:delta}
(n-1)-\frac{\sigma}{3}\leq\delta\leq n-1.
\end{equation}
In particular, the lower bound of $\delta$ in Eq.~\eqref{eq:delta} ensures that the adiabatic condition is violated, while the upper bound prevents $\kappa_\star$ from diverging. Notice also that, when the upper bound is saturated, $\delta=n-1$, we are in a special case: $\kappa_\star$ is not zero but it is discontinuous, namely $\lim_{r_+\to r_\star}\kappa_+>0$ while $\lim_{r_-\to r_\star}\kappa_-<0$.

Let us remark that Eq.~\eqref{eq:delta} is not the only constraint that $\phi(r,M)$ must satisfy in order for the scenario to be internally consistent.
Obviously the metric should remain nonsingular. There are two places in which the metric could become singular, at $r=r_\star$ and at $r=0$.
Computing the Ricci and Kretschmann scalar, one can check that, if Eq.~\eqref{eq:delta} holds only when $M=M_\star$, \emph{i.e.}, only when $r_\star$ is an horizon, then the metric is not singular. Additionally, it can be checked that, in order for the metric to not diverge at $r=0$, we must have that $e^{-\phi(r,M)}$ goes to a constant in the limit $r\rightarrow0$, while its derivative with respect to $r$ must vanish in this limit. Furthermore, we also want the instability timescale of the inner horizon to be much longer than the evaporation time, implying $\exp\left[-\phi(r_-,M)\right]\ll 1$. Strictly speaking, this condition is necessary only when the black hole is macroscopic or mesoscopic ($M\gg M_\star$). Indeed  $F'(r_-)$ approaches zero in the limit $M\to M_\star$, and, from Eq.~\eqref{eq:kappa:pm}, $\kappa_-$ might still be small for values of $\exp\left[-\phi(r_-,M)\right]$ of order unity or even greater.

If these properties are satisfied, the model can in principle be consistent from a geometric perspective. However, this limits the predictability of the late stages of evaporation, because one must resort to yet unknown physics beyond the effective description provided by regular black holes. Therefore, that the effective description breaks down just when it would become physically relevant, invalidates the original motivation and limits the scope of these models.

\section{Conclusion \label{sec:disc}}

We have analyzed two aspects of the physics of regular black holes, namely the instability of the inner horizon and the evaporation timescale. We have shown that the inner horizon is unstable on a finite timescale, and that this instability is a completely general feature of any model of regular black hole. On the other hand, we have also provided a self-consistent calculation of the evaporation time, showing that, in all the models currently presented in the literature, the complete evaporation cannot take place in a finite timescale. In combination with the unstable nature of the nonsingular core, the infinite evaporation time shows that these models are not self-consistent.

We have discussed that attaining finite evaporation times necessarily involves breaking either the adiabatic or the quasi-static conditions, which are necessary for the standard evaporation law to hold. This adds some degree of uncertainty. First, one has to decide the point at which these conditions break down: this additional information is just not encoded in the known models of regular black holes, which satisfy these conditions throughout their entire evolution. Moreover, one must also specify the details of how the evolution would continue. However, it seems that there is no general guiding principle restricting the freedom in describing the late stages of the evaporation beyond this critical point. In our opinion, this makes regular black holes much less appealing from a physical perspective as, in order to be viable, these models must fail precisely in the regime in which they should provide answers.

\acknowledgments

This publication was made possible through the support of the grant from the John Templeton Foundation No.51876. The opinions expressed in this publication are those of the authors and do not necessarily reflect the views of the John Templeton Foundation. Matt Visser was supported by the Marsden Fund, which is administered by the Royal Society of New Zealand. Matt Visser would like to thank SISSA and INFN (Trieste) for hospitality during the early phase of this work.

\bibliography{refs}	

\end{document}